# Metalenses with polarization-independent adaptive nano-antennas

Jianchao Zhang[1], Haowen Liang[1,2,*], Yong Long[1], Yongle Zhou[1], Qian Sun[1], Qinfei Wu[1], Xiao Fu[1,*], Emiliano R Martins[3], Thomas F Krauss[4], Juntao Li[1,*], Xue-Hua Wang[1]

[1] State Key Laboratory of Optoelectronic Materials and Technologies, School of Physics, Sun Yat-Sen University, Guangzhou 510275, China.

[2] Southern Marine Science and Engineering Guangdong Laboratory, (Zhuhai), Zhuhai 519080, China.

[3] São Carlos School of Engineering, Department of Electrical and Computer Engineering, University of São Paulo, 13566-590, Brazil.

[4] School of Physics, Engineering and Technology, University of York, York, YO10 5DD, UK.

E-mail: lianghw26@mail.sysu.edu.cn; xiaof58@mail.sysu.edu.cn; lijt3@mail.sysu.edu.cn

**Abstract:** Metalens research has made major advances in recent years. These advances rely on the simple design principle of arranging meta-atoms in regular arrays to create an arbitrary phase and polarization profile. Unfortunately, the concept of equally spaced meta-atoms reaches its limit for high deflection angles where the deflection efficiency decreases. The efficiency can be increased using nano-antennas with multiple elements, but their polarization sensitivity hinders their application in metalenses. Here, we show that by designing polarization-insensitive dimer nano-antennas and abandoning the principle of equally spaced unit cells, polarization-independent ultrahigh numerical aperture (NA=1.48) oil-immersion operation with an efficiency of 43% can be demonstrated. This represents a significant improvement on other polarization-independent designs at visible wavelength. We also use this single layer metalens to replace a conventional objective lens and demonstrate the confocal scanning microscopic imaging of a grating with a period of 300 nm at 532 nm operating wavelength. Overall, our results experimentally demonstrate a novel design concept that further improves metalens performance.

## 1. Introduction

Metalenses are constructed from sub-wavelength, equally-spaced meta-atoms that apply a phase and polarization distribution to the incident light field [1-3]; this distribution generates the desired wave front for imaging or other functions. While metalenses cannot completely replicate the properties of conventional objective lenses, they can reduce the cost, complexity and volume of compact and integrated imaging systems[4-14]. Furthermore, metalenses have already shown equal or superior properties for some select applications, such as optical trapping[15-17] or narrow-band wavelength confocal imaging[18-20].

The beauty of this simple design of placing meta-atoms into equally-spaced units unfortunately reaches its limit for high numerical aperture (NA) operation, because the discretized phase profile is less and less able to replicate the continuous profile, so the deflection efficiency decreases for high deflection angles[20,21]. Significant improvements can be obtained by using nano-antennas that achieve directional light scattering[19]. These nano-antennas consist of multiple elements, but they are polarization sensitive, which limits their practicality. In order to achieve high deflection efficiency also at high angles, it is therefore imperative to consider alternative design approaches. Here, we introduce a design that combines polarization-independent dimer nano-antennas placed into spatially varying unit cells, which we refer to as "adaptive" placement, on the periphery of the metalens with conventional unit cells at the center. The nano-antennas we use consist of two pillars in one unit cell, also known as "supercells"[22,23], or, more recently, "metagratings" [24,25], that vary with the radius of the metalens. The advantage of this approach is that it can achieve wave manipulation with a full $2\pi$ phase gradient by each nano-antenna functioning as a single Fresnel zone within State-of-the-Art fabrication constrains. Additionally, because of their circularly symmetric arrangement, the unit cell can be adjusted with increasing radius, thereby mapping ideally onto the size of the corresponding Fresnel zone. Thus, this new degree of freedom of adapting the nanoantenna placement to the specific phase requirement enables the design of high-efficiency metalenses for polarization-independent, high NA operation.

Some of the advantage of using nano-antennas with multiple elements ("metagratings") in the context of metasurfaces has already been described. Sell et al. [26] and Shi et al. [27] have demonstrated deflectors with large deflection angle and high deflection efficiency. Other researchers have achieved high NA focusing from the visible to the microwave regime [19,28]. The idea of using variable-size supercell unit cells for high NA metalens has also been introduced [29], but only by simulation. Furthermore, all the nano-antenna designs used so far are polarization-sensitive [19,28,29] and feature relatively low focus efficiency. Therefore, the polarization-independent adaptive nano-antenna approach we put forward here to achieve both high NA and high focusing efficiency has not been demonstrated yet in experiment, especially in the technologically important visible wavelength range (see also table 1).

We implement a number of important design improvements to achieve this high performance. Our experimental demonstration is based on crystalline silicon (c-Si) both for reasons of high refractive index and for its technological relevance. As we have shown before, c-Si, despite its intrinsic loss, is a very suitable material for realizing metalenses in the visible wavelength regime[18]. In particular, we adjust the phase of the supercell for every Fresnel zone, we use a fabrication-tolerant dimmer nano-antenna and we improve the design towards better polarization-insensitive operation. Together, these improvements lead to a significant improvement in overall performance. By immersion in cedar oil, which has a refractive index of 1.515, our metalens [Fig. 1(a)] experimentally achieves an NA of 1.48 at the wavelength of 532 nm. The highest focusing efficiency for polarization-independent operation is 43%, which increases slightly to 48% for operation with linearly polarized light; this is the highest efficiency for a polarization-independent, high NA metalens by a significant margin (see table 1).

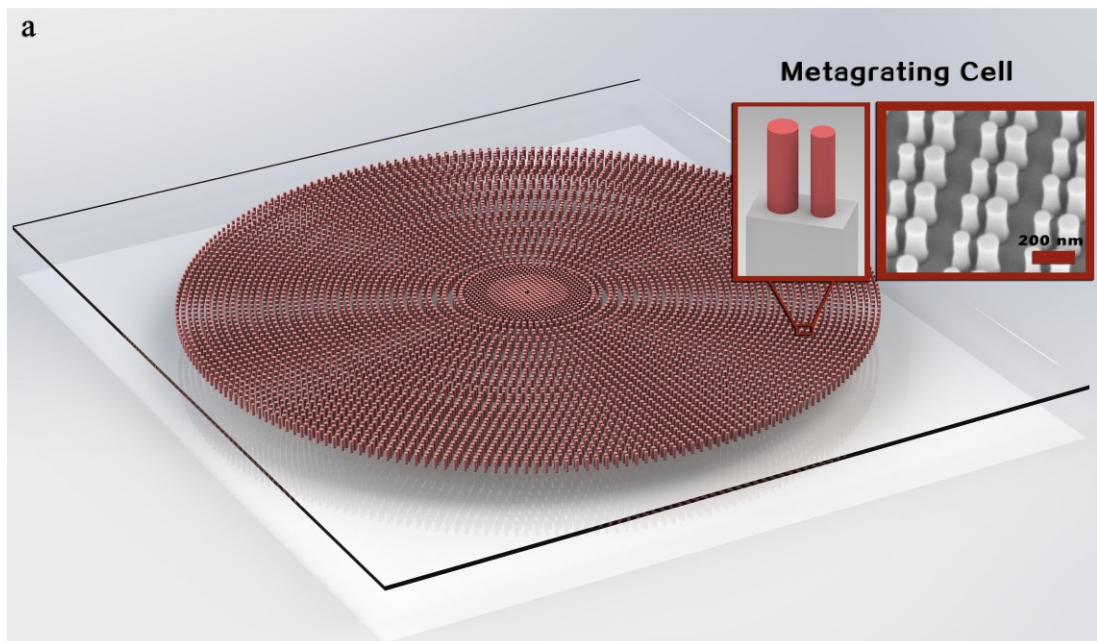

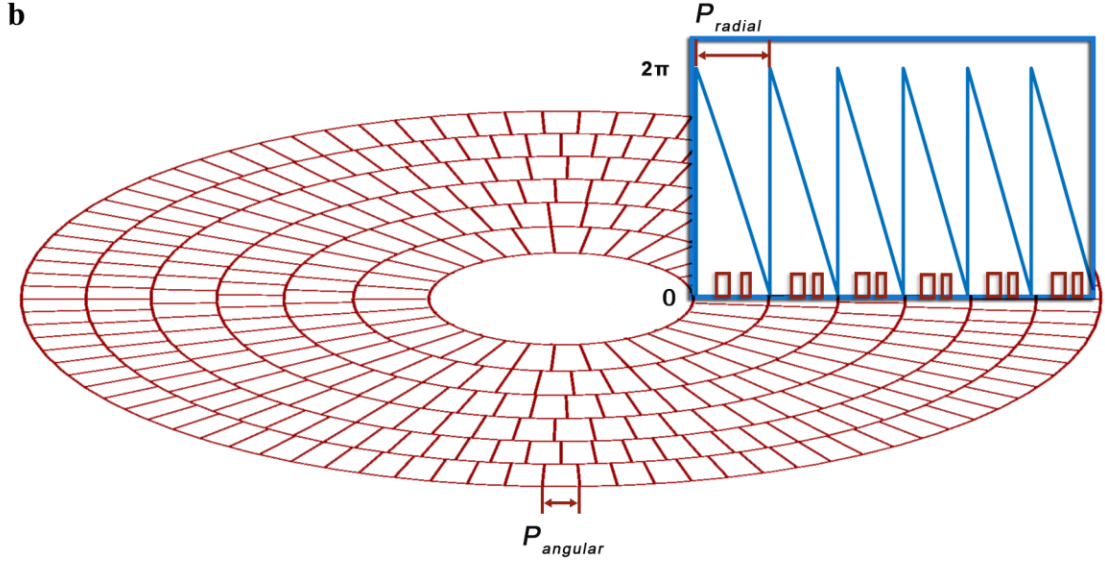

**Fig. 1. Schematic of the metalens. a** Sketch of the metalens. The inset shows a scanning electron microscopy (SEM) image of the dimmer nano-antennas. **b** Arrangement for the outer deflector. The phase profile (blue lines), corresponding to the Fresnel zones, is shown in the blue box.

## 2. Results

**Design of the metalens.** The phase profile of the metalens is encoded as a hyperbolic distribution,

$$\psi(r) = \frac{2\pi}{\lambda} n_g (f - \sqrt{r^2 + f^2}) \qquad (1)$$

where λ is the operating wavelength of 532 nm, $n_g$ is the background refractive index of 1.515, $f$ is the target focal length, and $r$ is the radial distance to the center.

In the large-angle deflection area, i.e. on the periphery of the lens, the phase profile can be approximated by a straight line with fixed phase gradient (see Fig. 1b). Therefore, every 2π period corresponds to a single deflection angle. Using this approximation, we adapt the dimer nano-antenna placement (Fig.1) to achieve a phase profile in each 2π period ($P_{radial}$ in Fig. 1b) that generates the required deflection angles, according to their radial positions. The diffraction period now depends on the radial distance covered by the 2π phase shift; therefore, $P_{radial}$ becomes a function of the metalens' radius[29]. In addition, we adjust the dimer parameters, i.e. the position of the two pillars, for every 1 degree in polar angle. Within 1 degree, we only change the $P_{radial}$ to match the 2π phase period. In the azimuthal dimension, the nano-antennas are placed according to their specific transverse period $P_{angular}$ in Fig.1b, filling the phase cycle rings with the same deflection angle. The difference to the conventional phase mapping method, where each meta-atom represents one phase, is that the dimer nano-antenna creates a smoother and more continuous phase gradient, thereby better approximating the phase distribution of an ideal lens.

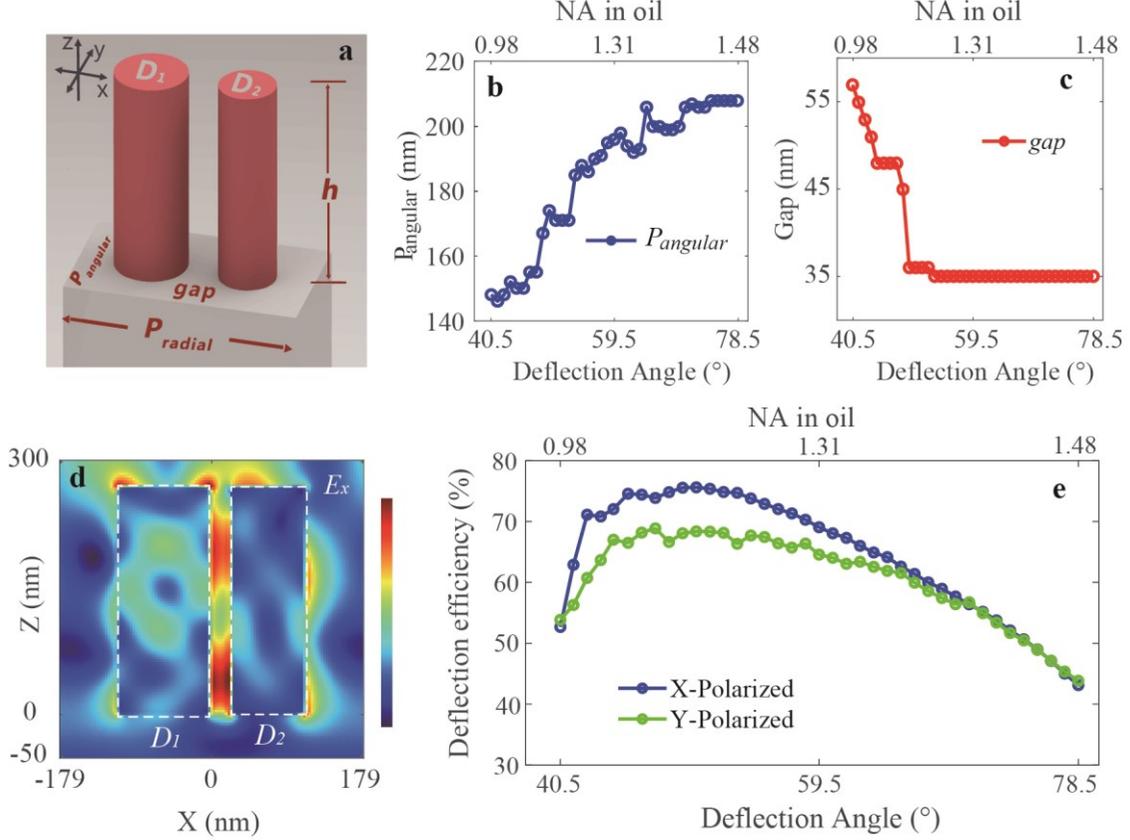

**Fig. 2 Nano-antenna parameters at 532 nm wavelength. a** Sketch of the unit cell, which consists of two circular pillars with a height of 270 nm. The directions of $P_{radial}$ and $P_{angular}$ are defined as x and y directions in the model. Optimized values of **b** $P_{angular}$ and **c** the *gap* between the two pillars for maximizing the deflection efficiency. **d** Simulated normalized electric field ($E_x$) profile along the radial direction ($P_{radual}$) for a dimer nano-antenna at the deflection angle of 78.5° with a *gap* of 35 nm and a $P_{augular}$ of 208 nm. The white frame indicates the location of the pillars. **e** Deflection efficiency of the nano-antenna with different deflecting angles for x-polarized and y-polarized beam.

The unit cell is comprised of two pillars separated by a *gap* (Fig. 2a). The period of these unit cells in the radial direction ($P_{radial}$) is determined by the grating equation:

$$P_{radial}(r) = \lambda / (n_g \sin \theta(r)) \tag{2}$$

Where *θ(r)* is the target deflection angle, which is a function of metalens' radius *r*. The parameters to optimize are the height, the *gap* between the pillars, the period in azimuthal direction ($P_{angular}$) and the diameters of the two pillars ($D_1$, $D_2$). The dimers exhibit an intrinsic polarization-dependence, because the unit cell is not rotationally symmetric. In order to minimize this polarization-dependence, we perform an adjoint-based optimization[30-34] (See Section 1 of supplemental for details). Using this method, the optimization, despite the number of parameters involved, converges rapidly and with high accuracy.

The efficiency of diffraction into the first order ($T_{+1}$) is set as the target function. The deflection efficiency is defined as the ratio of the transmitted light into $T_{+1}$ level to the total incident light. We aim to achieve a high and equal deflection efficiency for two orthogonal linearly polarizations in the optimization.

The optimized parameters are shown in Fig. 2b-2c. The height of the c-Si circular pillars is optimized to be 270 nm. Considering the feasibility of the fabrication process, we set the lower limit of the *gap* to 35 nm and the lower limit of the $P_{angular}$ to 145 nm. The parameter optimization of $P_{angular}$ and of the *gap* is influenced by the target deflection angle. As the deflection angle increases, the optimum *gap* reduces from 57 nm to 35 nm, and the optimum $P_{angular}$ increases from 145 nm to 208 nm, respectively (Fig. 2b-c). The diameters of the dimer can be fixed to a certain combination ($D_1$ = 110 nm, $D_2$ = 88 nm) while maintaining the high deflection efficiency (Fig. S1b- S1c). Thus, for the entire structure, the sizes of the two pillars remain fixed, and the desired phase profile is achieved entirely by changing the position of these two pillars. As it is easier to change the structures' positions than to change the structures' sizes in fabrication[35], this approach has the technological advantage of leading to higher experimental focusing efficiency. Furthermore, as shown in Fig. 2d, the electric field intensity is mainly concentrated in the *gap* between the pillars, which leads to reduced absorption losses[36].

For a metalens with an NA of 1.48 (in oil), the surface area that covers the deflection angles between 40.5° to 78.5°, which corresponds to NA values between 0.98 to 1.48, accounts for 97% of the total metalens area. The optimized deflection efficiency as a function of this range of deflection angle is shown in Fig. 2e. For the surface area corresponding to a large deflection angle between 65.5° to 78.5°, which accounts for 81% of the total metalens area, the difference in deflection efficiency between two orthogonal linearly polarizations is less than 1.5%. We note that similar nano-antennas have been reported in the literature[19,36], but with a much higher efficiency difference between the two polarizations.

The area below 40.5°, in the center of the lens, is implemented using the method of phase mapping, which better matches the desired phase profile at low angles[19,28,37] (Specific parameters are shown in Section 2 of supplemental). Based on our simulation, this boundary between the phase mapping and adaptive design achieves the highest focusing efficiency overall (see Section 3 of supplemental for details).

**Fabrication and characterization.** The metalens is fabricated on a c-Si on sapphire (SOS) wafer (see "**Methods**" for fabrication details). The radius of the fabricated metalens is 100 µm, which is sufficient for the confocal microscope imaging application described next. We note that there is no limit to the overall area of the metalens in our design. Hence, the design is easily scalable towards larger areas. SEM (Zeiss Auriga) micrographs of the fabricated metalens are shown in Fig. 3. Note the high quality of the fabricated structures, which explains the close agreement between simulation and experiment.

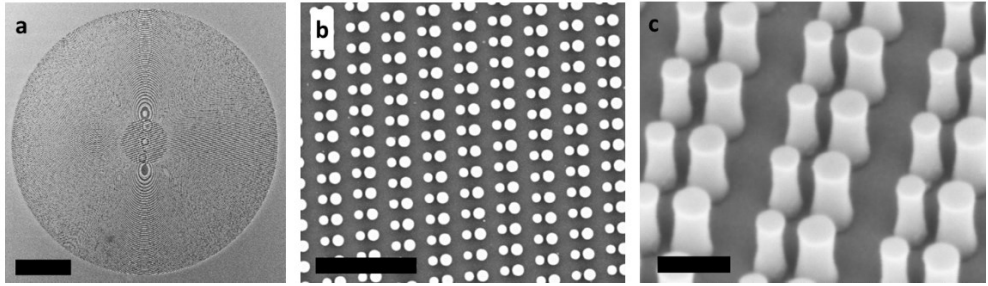

**Fig. 3. SEM images of the metalens. a** Top view of the complete device. **b** Top and **c** side view (30°) on the periphery of the metalens. Scalebars: **a** 40 µm, **b** 1 µm, and **c** 200 nm.

The experimental focusing performance at 532 nm wavelength is characterized by a set of microscopic imaging systems (See Section 4 of supplemental and "**Methods**" for the details of optical setup). The measured results are shown in Fig. 4. The focal length is 20.40 µm, corresponding to an ultrahigh NA of 1.48 for the radius of 100 µm and the index of 1.515. The full width at half maximum (FWHM) size of the focal spot in the horizontal and vertical directions are 268 nm (0.51 λ) and 209 nm (0.39 λ) for horizontal polarized light (Fig. 4a). Here, the FWHM was obtained from an Airy fitting to the one-dimensional distribution along the corresponding direction. The FWHM for unpolarized incident beam is calculated to be 238 nm, by averaging the point spread function (PSF) of the linearly polarized incident beam with different polarization angles (Fig. 4b).

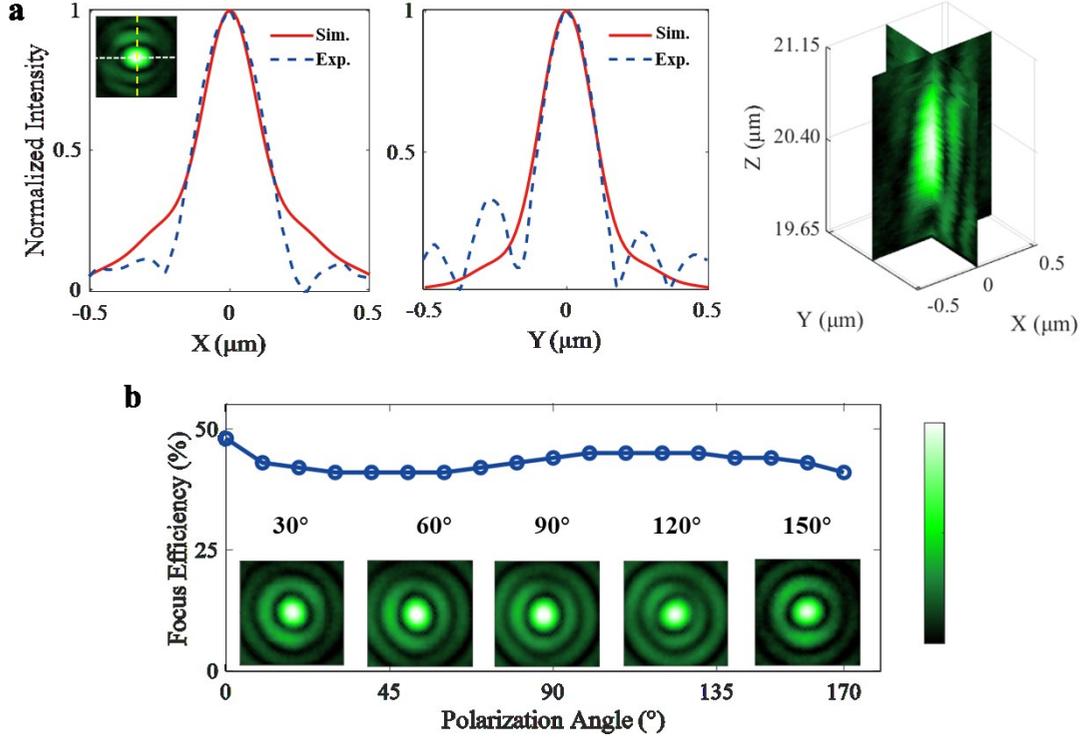

**Fig. 4. Experimental focusing performance of the metalens at 532 nm wavelength. a** PSF for horizontal (left) and vertical (right) polarization, comparing experiment (blue dotted line) with simulation (red solid line). The inset figure on the upper left is the experimentally obtained PSF in both directions. The three-dimensional image on the right fully describes the experimental focal spot. **b** Measured focusing efficiency and PSF for linearly polarized light with different polarization angles.

We define the focusing efficiency as the ratio between the transmission power in the aperture of the focal spot. As shown in Fig. 4a, the focusing efficiency reaches 48% for horizontal polarized light, which is obtained by measuring the power within an circle of diameter 3× FWHM of the spot in transmission. The focusing efficiency for unpolarized incident beam is calculated to be 43%, by averaging the focusing efficiency of the linearly polarized incident beam with different polarization angles Fig. 4b. This result confirms the polarization-insensitive characteristics of our metalens.

To confirm the experiment results, we perform finite-difference time-domain (FDTD) simulations to simulate a metalens with a diameter of 55.00 μm and a horizontal linearly polarized (polarization angle of 0°) incidence beam by the commercial software FDTD Solutions (Lumerical Inc.) (See Section 5 of supplemental for details). As shown in Fig. 4a, the PSF in simulations are consistent with the corresponding PSF results in experiment (See Section 5 of supplemental for the simulation methods).

**Confocal imaging.** Finally, we demonstrate that our novel metalens can be used as the objective lens in a high-resolution confocal imaging system. We placed the lens into a commercial laser scanning confocal microscope (WITec alpha300S) using a 532 nm diffraction-limited laser beam from a single mode fiber as the light source. A metallic grating was used as the imaging object, featuring 300 nm period and 150 nm lines (50% fill-factor) in a 100 nm thick gold film. With the small focal spot size of our lens, it is clear that the shape of the grating can be accurately identified, with good contrast (Fig.5). The grating period corresponds to 0.564 λ, which represents the highest confocal imaging resolution reported for a metalens in the literature to our knowledge[18-20]. This exceptional performance demonstrates the potential for a single layer metalens to replace a conventional objective lens even for the demanding confocal scanning microscopic imaging application.

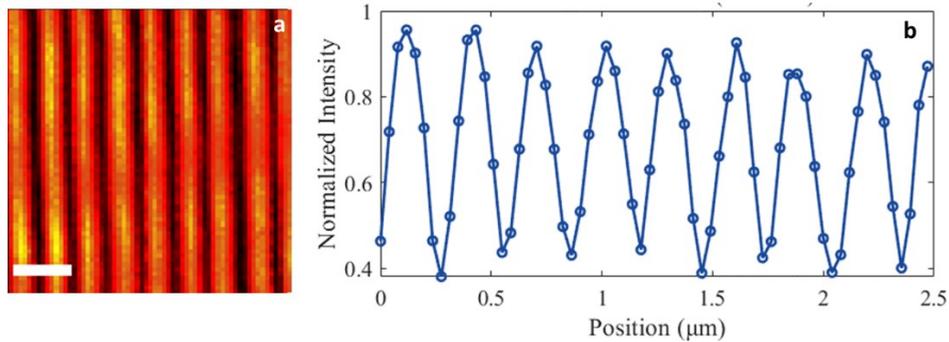

**Fig. 5. Confocal imaging results. a** Experimental confocal scanning microscopic image using the immersion metalens with NA=1.48 of metallic gratings with a period of 300 nm and a line width of 150 nm. Scale bar: 500 nm. **b** The intensity along the horizontal direction through the center of **a**. Mean peak-to-peak value of (a) is 294 nm.

3. Discussion

It is interesting to compare the efficiency and performance metrics of different high NA metalenses, which are reported in Table 1. We note that most of them operate in the red and NIR wavebands, which makes it easier to achieve good performance, while we operate at the more demanding 532 nm wavelength. We also include some polarization-dependent designs at high NA, but they require polarization-controlled light sources. For the benefit of the reader and to enable a fair comparison, we have limited table 1 to lenses operating at visible or near-visible wavelengths with an NA close to or above 1, because this is the most demanding regime for high efficiency operation. Taking all these considerations into account, we note that our metalens is competitive with the best polarization-dependent design in the visible (e.g. PB phase-based metalens in Ref[18,20]), and it is much better than any other polarization-independent design.

| Table 1 Summary of Previously Reported Experimental High-NA Metalenses. | | | | | | | |
|---|---|---|---|---|---|---|---|
| Reference | Materials | Polarization | NA | Radius (μm) | Wave-length (nm) | Focusing Efficiency (%) | Aperture Size in Focusing Efficiency Calculation |
| This work | c-Si | Polarization-independent | 1.48 n=1.515 | 100 | 532 | 43 | 3×FWHM |
| | | | | | | 59 | 13.4×FWHM |
| Mansouree et al.[34] (Overall optimization) | a-Si | Polarization-independent | 0.94 | 25 | 850 | 49 | 13.4×FWHM |
| Paniagua-Domínguez et al.[19] (Metagrating) | a-Si | Polarization-independent | 0.99 | 200 | 715 | 10 | 5.2×FWHM [a] |
| Hail et al.[38] (Phase matching) | a-Si | Circular polarization | 1.4 n=1.52 | 116 | 820 | 21 | 3.8×FWHM [a] |
| Chen et al.[20] (Phase matching) | TiO2 | Circular polarization | 1.1 n=1.52 | 225 | 532 | 52 | Undefined |
| Liang et al.[18] (Phase matching) | c-Si | Circular polarization | 1.48 n=1.51 | 500 | 532 | 48 | 3×FWHM |
| [a] A square aperture was used to collect the power of the spot light. The half side length of aperture was set to be the value of aperture size in the table. | | | | | | | |

## 4. Conclusion

We demonstrate a metalens based on polarization-independent adaptively placed nano-antennas with ultrahigh NA of 1.48, a focusing efficiency of above 43% and reliable high resolution confocal scanning microscopic imaging ability in oil immersion for an unpolarized incident beam at the wavelength of 532 nm.

Our metalens achieves efficient polarization insensitive deflection at high angles, which is the key novelty reported here. Furthermore, the phase of the unit cell for every Fresnel zone has been carefully adjusted; this means that we break with the fundamental design principle of most metalenses, whereby the size of the unit cell is kept constant, yet we show that this deviation from the standard design rules is beneficial. We show a number of subtle, but very important improvements that lead to this significant improvement in performance. We use a more fabrication-tolerant unit cell by keeping the size of the dielectric pillars that define the unit cell constant, while only changing their positions. We also note that our dimer nano-antenna design favours light confinement between the pillars rather than inside them, which we have previously shown to be beneficial for low-loss operation[36]. Finally, we improved the design towards better polarization-insensitive operation by achieving nano-antennas with high and nearly equal deflection efficiency for two orthogonal linearly polarizations via an adjoint-based optimization.

**Methods**

**Fabrication.** The metalens is fabricated on a c-Si on sapphire (SOS) wafer. The c-Si film is thinned down to a thickness of 270 nm. The pattern is defined in negative resist (hydrogen silsesquioxane, HSQ, 200 nm thick film, Dow Corning) by electron beam lithography (EBL, Raith Vistec EBPG-5000plusES). Then the pattern is transferred into the c-Si film by Inductively Coupled Plasma (ICP, Oxford Instruments PlasmaPro 100ICP180) etching using HBr chemistry. Finally, the remaining HSQ is removed using hydrogen fluoride (HF).

**Measurement.** The experimental focusing performance is characterized by a microscopic imaging system, including a 100× oil immersion objective lens (Nikon CFI APOCHROMAT TIRF MRD01991), a tube lens (LBTEK MCX10619-A), a laser light source at 532 nm wavelength and a CCD camera (Kiralux CS895CU) (See Section 4 of supplemental for details). The tube lens has a focal length of 30 cm, which leads to a 150× system magnification. The actual size of focal light intensity distribution from the recorded CCD image can be calculated by dividing this system magnification into the size of the CCD.

**Acknowledgements**

This work was supported by the Guangdong Provincial Key R&D Program (No. 2019B010152001), National Natural Science Foundation of China (Nos. 11974436, 12074444, and 11704421), Guangdong Basic and Applied Basic Research Foundation (Nos. 2020B1515020019 and 2020A1515011184). H.L. acknowledges support by Innovation Group Project of Southern Marine Science and Engineering Guangdong. E.R.M. acknowledges support by Grant No. 2020/00619-4, São Paulo Research Foundation (FAPESP). T.F.K. acknowledges support by UK Research & Innovation under Contract Nos. EP/P030017/1 and EP/T020008/1.

**References**

1   Arbabi, E., Arbabi, A., Kamali, S. M., Horie, Y. & Faraon, A. Multiwavelength polarization-insensitive lenses based on dielectric metasurfaces with meta-molecules. *Optica* **3**, doi:10.1364/optica.3.000628 (2016).

2   Khorasaninejad, M. *et al.* Polarization-Insensitive Metalenses at Visible Wavelengths. *Nano Lett* **16**, 7229-7234, doi:10.1021/acs.nanolett.6b03626 (2016).

3   Arbabi, A., Horie, Y., Bagheri, M. & Faraon, A. Dielectric metasurfaces for complete control of phase and polarization with subwavelength spatial resolution and high transmission. *Nat Nanotechnol* **10**, 937-943, doi:10.1038/nnano.2015.186 (2015).

4   Khorasaninejad M, C. W., Devlin RC, Oh J, Zhu AY, Capasso F. . Metalenses at visible wavelengths: Diffraction-limited focusing and subwavelength resolution imaging. *Science* **352**, 1190-1194 (2016).

5   Li, B., Piyawattanametha, W. & Qiu, Z. Metalens-Based Miniaturized Optical Systems. *Micromachines (Basel)* **10**, doi:10.3390/mi10050310 (2019).

6   Pahlevaninezhad, H. *et al.* Nano-optic endoscope for high-resolution optical coherence tomography in vivo. *Nat Photonics* **12**, 540-547, doi:10.1038/s41566-018-0224-2 (2018).

7   Martins, A. *et al.* On Metalenses with Arbitrarily Wide Field of View. *ACS Photonics* **7**, 2073-2079, doi:10.1021/acsphotonics.0c00479 (2020).

8   Arbabi, E. *et al.* Two-Photon Microscopy with a Double-Wavelength Metasurface Objective Lens. *Nano Lett* **18**, 4943-4948, doi:10.1021/acs.nanolett.8b01737 (2018).

9   Arbabi, A. *et al.* Miniature optical planar camera based on a wide-angle metasurface doublet corrected for monochromatic aberrations. *Nat Commun* **7**, 13682, doi:10.1038/ncomms13682 (2016).

10  Afridi, A. *et al.* Electrically Driven Varifocal Silicon Metalens. *ACS Photonics* **5**, 4497-4503, doi:10.1021/acsphotonics.8b00948 (2018).

11  Aiello, M. D. *et al.* Achromatic Varifocal Metalens for the Visible Spectrum. *ACS Photonics* **6**, 2432-2440, doi:10.1021/acsphotonics.9b00523 (2019).

12  Chen, C. *et al.* Spectral tomographic imaging with aplanatic metalens. *Light Sci Appl* **8**, 99, doi:10.1038/s41377-019-0208-0 (2019).

13  Shalaginov, M. Y. *et al.* Single-Element Diffraction-Limited Fisheye Metalens. *Nano Lett* **20**, 7429-7437, doi:10.1021/acs.nanolett.0c02783 (2020).

14  Bosch, M. *et al.* Electrically Actuated Varifocal Lens Based on Liquid-Crystal-Embedded Dielectric Metasurfaces. *Nano Lett* **21**, 3849-3856, doi:10.1021/acs.nanolett.1c00356 (2021).

15  Markovich, H., Shishkin, II, Hendler, N. & Ginzburg, P. Optical Manipulation along an Optical Axis with a Polarization Sensitive Meta-Lens. *Nano Lett* **18**, 5024-5029, doi:10.1021/acs.nanolett.8b01844 (2018).

16  Chantakit, T. *et al.* All-dielectric silicon metalens for two-dimensional particle manipulation in optical tweezers. *Photonics Research* **8**, doi:10.1364/prj.389200 (2020).

17  Tkachenko, G. *et al.* Optical trapping with planar silicon metalenses. *Opt Lett* **43**, 3224-3227, doi:10.1364/OL.43.003224 (2018).

18      Liang, H. *et al.* Ultrahigh Numerical Aperture Metalens at Visible Wavelengths. *Nano Lett* **18**, 4460-4466, doi:10.1021/acs.nanolett.8b01570 (2018).

19      Paniagua-Dominguez, R. *et al.* A Metalens with a Near-Unity Numerical Aperture. *Nano Lett* **18**, 2124-2132, doi:10.1021/acs.nanolett.8b00368 (2018).

20      Chen, W. T. *et al.* Immersion Meta-Lenses at Visible Wavelengths for Nanoscale Imaging. *Nano Lett* **17**, 3188-3194, doi:10.1021/acs.nanolett.7b00717 (2017).

21      Arbabi, A., Horie, Y., Ball, A. J., Bagheri, M. & Faraon, A. Subwavelength-thick lenses with high numerical apertures and large efficiency based on high-contrast transmitarrays. *Nat Commun* **6**, 7069, doi:10.1038/ncomms8069 (2015).

22      Martins, E. R., Li, J., Liu, Y., Zhou, J. & Krauss, T. F. Engineering gratings for light trapping in photovoltaics: The supercell concept. *Physical Review B* **86**, doi:10.1103/PhysRevB.86.041404 (2012).

23      Martins, E. R. *et al.* Deterministic quasi-random nanostructures for photon control. *Nat Commun* **4**, 2665, doi:10.1038/ncomms3665 (2013).

24      Ra'di, Y., Sounas, D. L. & Alu, A. Metagratings: Beyond the Limits of Graded Metasurfaces for Wave Front Control. *Phys Rev Lett* **119**, 067404, doi:10.1103/PhysRevLett.119.067404 (2017).

25      Epstein, A. & Rabinovich, O. Unveiling the Properties of Metagratings via a Detailed Analytical Model for Synthesis and Analysis. *Physical Review Applied* **8**, doi:10.1103/PhysRevApplied.8.054037 (2017).

26      Sell, D., Yang, J., Doshay, S., Yang, R. & Fan, J. A. Large-Angle, Multifunctional Metagratings Based on Freeform Multimode Geometries. *Nano Lett* **17**, 3752-3757, doi:10.1021/acs.nanolett.7b01082 (2017).

27      Shi, T. *et al.* All-Dielectric Kissing-Dimer Metagratings for Asymmetric High Diffraction. *Advanced Optical Materials* **7**, doi:10.1002/adom.201901389 (2019).

28      Kang, M., Ra'di, Y., Farfan, D. & Alù, A. Efficient Focusing with Large Numerical Aperture Using a Hybrid Metalens. *Physical Review Applied* **13**, doi:10.1103/PhysRevApplied.13.044016 (2020).

29      Byrnes, S. J., Lenef, A., Aieta, F. & Capasso, F. Designing large, high-efficiency, high-numerical-aperture, transmissive meta-lenses for visible light. *Opt Express* **24**, 5110-5124, doi:10.1364/OE.24.005110 (2016).

30      Wen, F., Jiang, J. & Fan, J. A. Robust Freeform Metasurface Design Based on Progressively Growing Generative Networks. *ACS Photonics* **7**, 2098-2104, doi:10.1021/acsphotonics.0c00539 (2020).

31      Chung, H. & Miller, O. D. High-NA achromatic metalenses by inverse design. *Opt Express* **28**, 6945-6965, doi:10.1364/OE.385440 (2020).

32      Backer, A. S. Computational inverse design for cascaded systems of metasurface optics. *Opt Express* **27**, 30308-30331, doi:10.1364/OE.27.030308 (2019).

33      Chung, H. & Miller, O. D. Tunable Metasurface Inverse Design for 80% Switching Efficiencies and 144° Angular Deflection. *ACS Photonics* **7**, 2236-2243, doi:10.1021/acsphotonics.0c00787 (2020).

34      Mansouree, M., McClung, A., Samudrala, S. & Arbabi, A. Large-Scale Parametrized

Metasurface Design Using Adjoint Optimization. *ACS Photonics* **8**, 455-463, doi:10.1021/acsphotonics.0c01058 (2021).

35  Li J, W. T., O'Faolain L, Gomez-Iglesias A, Krauss TF. . Systematic design of flat band slow light in photonic crystal waveguides *Opt Express* **16**, 6227-6232, doi:10.1364/oe.16.006227 (2008).

36  Sun, Q. *et al.* Highly Efficient Air-Mode Silicon Metasurfaces for Visible Light Operation Embedded in a Protective Silica Layer. *Advanced Optical Materials* **9**, doi:10.1002/adom.202002209 (2021).

37  Zhou, Z. *et al.* Efficient Silicon Metasurfaces for Visible Light. *ACS Photonics* **4**, 544-551, doi:10.1021/acsphotonics.6b00740 (2017).

38  Hail, C. U., Poulikakos, D. & Eghlidi, H. High-Efficiency, Extreme-Numerical-Aperture Metasurfaces Based on Partial Control of the Phase of Light. *Advanced Optical Materials* **6**, doi:10.1002/adom.201800852 (2018).

# Supplementary Information

# Metalenses with polarization-independent adaptive nano-antennas

Jianchao Zhang[1], Haowen Liang[1,2,*], Yong Long[1], Yongle Zhou[1], Qian Sun[1], Qinfei Wu[1], Xiao Fu[1,*], Emiliano R Martins[3], Thomas F Krauss[4], Juntao Li[1,*], Xue-Hua Wang[1]

[1] State Key Laboratory of Optoelectronic Materials and Technologies, School of Physics, Sun Yat-Sen University, Guangzhou 510275, China.

[2] Southern Marine Science and Engineering Guangdong Laboratory, (Zhuhai), Zhuhai 519080, China.

[3] São Carlos School of Engineering, Department of Electrical and Computer Engineering, University of São Paulo, 13566-590, Brazil.

[4] University of York, Department of Physics, York YO10 5DD, UK.

E-mail: lianghw26@mail.sysu.edu.cn; xiaof58@mail.sysu.edu.cn; lijt3@mail.sysu.edu.cn

## 1. Adjoint optimization algorithm.

The flow of the adjoint optimization algorithm is shown in Fig. S1(a). Firstly, we enter random initial parameters for the geometric size combination of the nano-antenna cell. Then we use finite-difference time-domain (FDTD) simulations from the commercial software FDTD Solutions (Lumerical Inc.) to carry out the forward simulation, where the incident uniform plane waves of TE and TM are transmitted upward in the nano-antenna substrate. The deflection efficiency of the nano-antenna under these parameters can be calculated from the far-field results of the monitor above the nano-antenna, and the three-dimensional electric field distribution $E_F$ in the vicinity of the dimer region can be obtained. The next step is to determine whether the calculation has converged. If the deflection efficiency is stable or the number of repetitions reaches the set value, the cycle can be terminated. If not, then we enter the next step of adjoint simulation by using a surface light source with ideal phase gradient above the dimer nano-antenna to illuminate the dimer downward to obtain the three-dimensional electric field distribution $E_A$ near the dimer region. By making an overlapping integration of $E_A$ and $E_F$, we can calculate the gradient required to achieve the ideal phase plane for each geometric parameter. After the judgment, the gradient is applied to the structure size for the next cycle until the result converges. By entering the initial value combination several times, the global optimal parameter combination can be found.

To verify the accuracy, we perform parameter sweepings of the dimer diameters, *gap* and $P_{angular}$ at the deflection angles of 78.5° and 50.6° for circularly polarized light (Fig. S1b-S1c). The deflection efficiency distribution is in accordance with the algorithm optimization result, which means the result obtained by the algorithm is the global optimal solution.

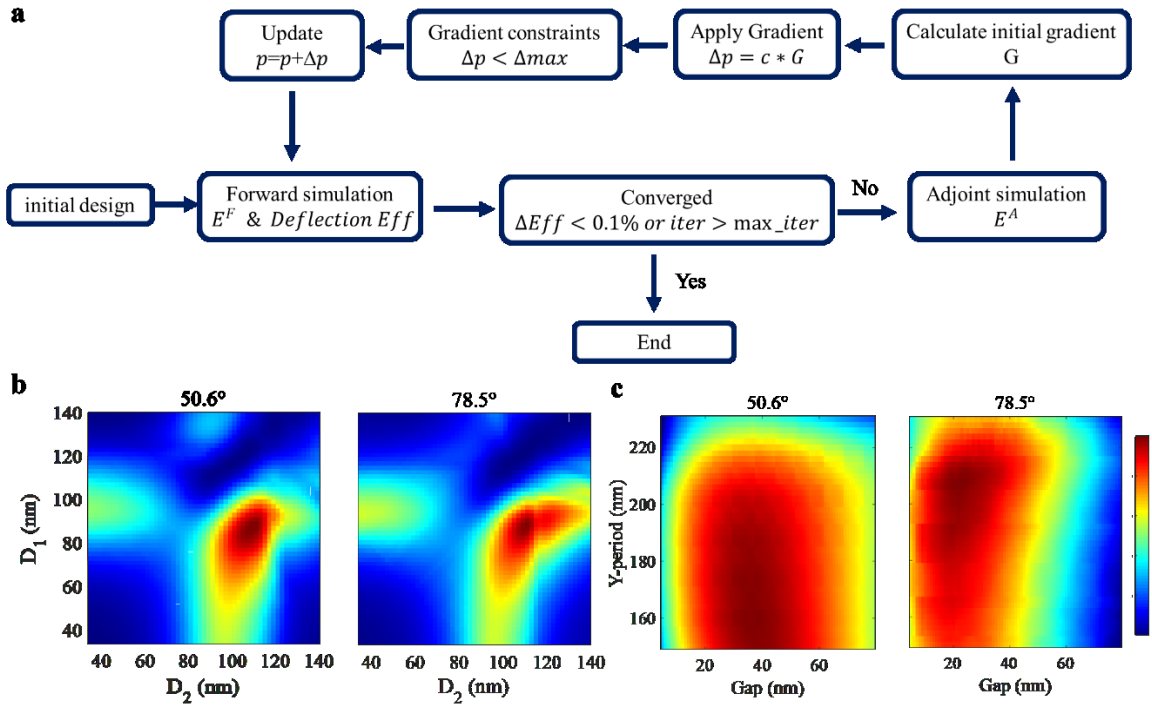

**Fig. S1. Adjoint optimization algorithm graph and parameter sweep results. a** Adjoint optimization algorithm graph. **b** Parameter sweepings of the dimer diameters with the height of 270 nm at the deflection angles of 50.6°and 78.5°. The deflection efficiency distribution is related to the diameters. $P_{angular}$ and gap are set to the optimized lengths, 171 nm and 36 nm for 50.6°, and 208 nm and 35 nm for 78.5°. **c** $P_{arameter}$ sweepings of the gap and $P_{angular}$ for the two deflection angles. The dimer diameters are set to 110 nm and 88 nm.

## 2. Phase mapping unit cell parameters.

We perform FDTD simulations to determine the meta-atom parameters in the phase mapping region, i.e. in the centre of the lens. We then perform a parameter sweep of the unit cell length (a) and the duty ratio (pillar diameter/unit-cell size) (Fig. S2a-S2b). The phase control of the pillars has the same baseline, assuming empty cell (0 duty ratio) bringing 0 additional phase.

We choose circular silicon pillars with a = 180 nm due to their high transmittance. Correspondingly, we select 6 equidistant low-loss parameters in the black line to cover the full 2π phase control uniformly. As shown in Table S1, there are 6 circular pillars (including 0 duty ratio) in 0~2π with the phase difference of 1/3π, the lowest transmittance of the pillars is above 65%.

| Table S1. Feature Size of Pillars with 180 nm Unit Cell Size and 270 nm Height | | | | | | |
|---|---|---|---|---|---|---|
| Phase (rad) | 0 | π/3 | 2π/3 | π | 4π/3 | 5π/3 |
| D (nm) | 0 | 74 | 90 | 100 | 112 | 126 |

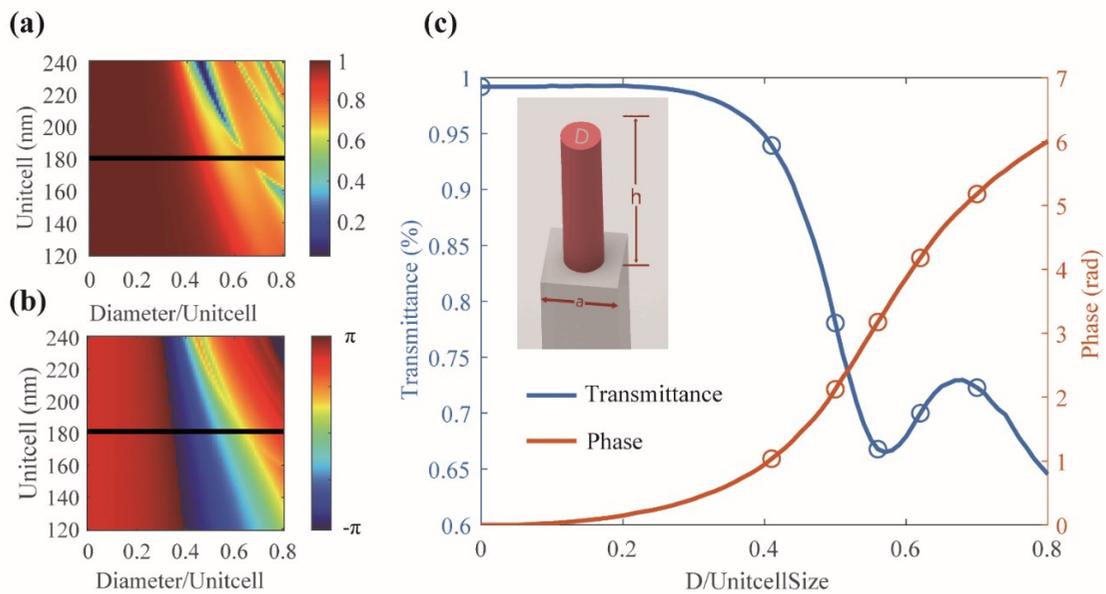

**Fig. S2. Phase mapping pillar parameters.** Calculation of **a** the transmittance and **b** the phase of the periodic circular pillars. **c** Transmittance and phase of the selected 6 pillars.

### 3. Nano-antennas arrangement method along radial dimension

To determine the boundary between the phase mapping and nano-antenna parts of the lens, we simulated a series hybrid metalens with the same NA of 1.48 and a diameter of 55.00 μm using different boundaries between the two components by FDTD and calculating their focus efficiency (Fig. S3a). We note that for a boundary around NA≈1, the focusing efficiency reaches 58% and remains constant towards smaller NAs. We chose the boundary at NA = 0.98, where the hybrid metalens achieves the best focus efficiency of 59%. Note that metalenses with different boundaries and the same NA have focal spots with an approximately equal size and shape characteristic (Fig. S3b).

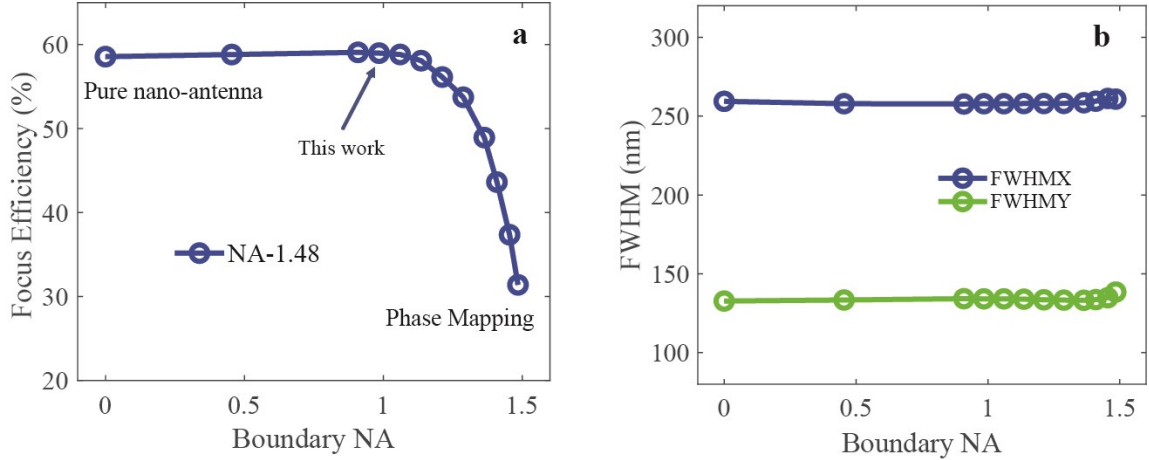

**Fig. S3. Comparison between phase mapping metalens and nano-antennas metalens. a** Simulated focusing efficiency of metalenses with different boundary NA between phase mapping section and nano-antenna section and the same NA of 1.48. **b** Corresponding spot size of the metalenses with different boundary NA.

## 4. Experimental setup for the measurement

As shown in Fig. S4, the experimental focusing performance is characterized by a microscopic imaging system, including a 100× oil immersion objective lens (Nikon CFI APOCHROMAT TIRF MRD01991), a tube lens (LBTEK MCX10619-A), a laser light source at 532 nm wavelength and a CCD camera (Kiralux CS895CU). The tube lens has a focal length of 30 cm, which leads to a 150× system magnification. The actual size of focal light intensity distribution from the recorded CCD image can be calculated by dividing this system magnification into the size of the CCD.

The linear polarizer (LP) is used to generate linearly polarized light, and then the polarization direction of the linearly polarized light can be adjusted by rotating the half-wave plate (HWP). In addition, we add a blank sapphire chip with a crystal orientation orthogonal to the substrate in front of the substrate to compensate the birefringence effect of the sapphire substrate. The designed NA of 1.48 of the metalens is close to the NA of 1.49 of the microscope oil immersion objective lens. This microscopic imaging system can measure the enlarged light spot but cannot detect the $I_z$ component[1].

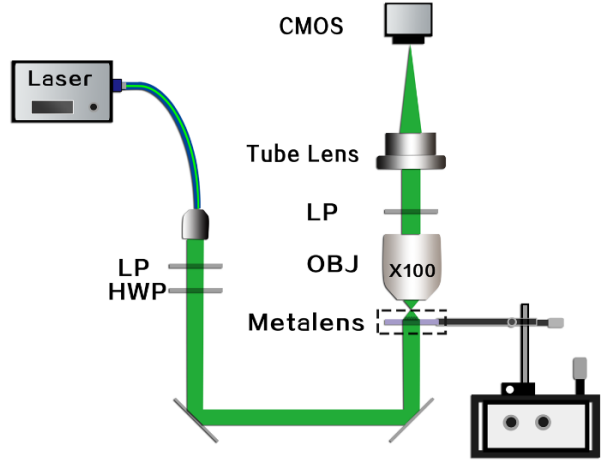

**Fig. S4. Schematic diagram of the metalens focal spot measurement.**

## 5. Simulation of the metalens performance.

We firstly simulate a metalens with a diameter of 55.00 μm and a horizontal linearly polarized by using the FDTD simulation. The simulated focal length is 5.49 μm, corresponding to the ultrahigh NA of 1.48 in immersion oil (Fig. S5c). Because the metalens itself is rotationally symmetric, its focusing efficiency for a linearly polarized and for an unpolarized incident beam should be the same, which is calculated to be 59%.

As shown in Fig. S5b, for this ultrahigh NA metalens, the FWHM of the focal spot in the horizontal and vertical directions are calculated to be 258 nm (0.485 λ) and 133 nm (0.249 λ), respectively. We note that the focal spot is not an ideal Airy pattern, as the linearly polarized beam is focused into a flat elliptical spot. This is a well-known distortion for a high-NA objective, and it is a consequence of the depolarization effect[2].

The $I_z$ component occupies more than 40% of the energy of the focal spot (Table S2). For the linearly polarized focal spot, the dominant proportion of $I_z$ component makes the overall spot close to the spindle shape. A similar experimental observation of the spindle-shaped focal spot of a super-focusing metalens has been reported in the literature[3].

To better characterize the spot, we further obtain the point spread function (PSF) of an idea lens with NA = 1.48 by using the Richards-Wolf vector diffraction integration method (VDIM). The calculated PSF is slightly flatter than the FDTD result. The basic reason is that the nano-antenna cell could not have the same deflection effect on x- and y-polarization at all angles completely (Fig. 2e), accordingly caused the higher $I_z$ component.

Finally, we also simulate the PSF in the experiment by convolving the PSF with $I_{xy}$ components of the metalens and the PSF with $I_{xy}$ components of the experiment microscope imaging system (Fig. S4)[4]. The PSF in simulations is consistent with the experimentally measured result (Fig. S5d).

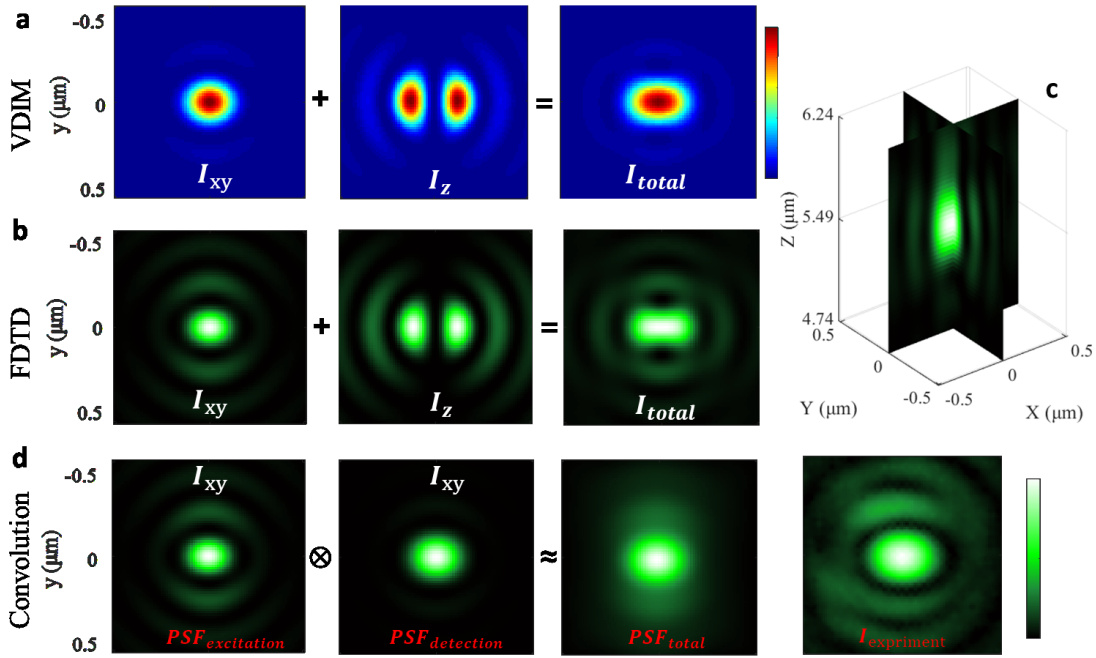

**Fig. S5. Simulated focusing performance of metalens by a horizontally linearly polarized incident beam. a** PSF of an idea lens calculated through VDIM. **b** PSF of the metalens by FDTD simulation. **c** The corresponding three-dimensional image of the focal spot by FDTD simulation. **d** from left to right: the PSF with $I_{xy}$ components of the metalens by FDTD simulation, the PSF with $I_{xy}$ components of the experiment microscope imaging system (Fig. S4) by VDIM calculation, the convolution result of the first two PSFs, and the PSF in the experiment.

| Table S2. The energy ratio by FDTD and VDIM. | | | |
|---|---|---|---|
| Methods | FWHM (nm) | $I_{xy}$ (%) | $I_z$ (%) |
| FDTD | 260/134 | 59 | 41 |
| VDIM | 250/169 | 74 | 26 |

**Reference**

1  Liu, T., Yang, S. & Jiang, Z. Electromagnetic exploration of far-field super-focusing nanostructured metasurfaces. *Opt Express* **24**, 16297-16308, doi:10.1364/OE.24.016297 (2016).
2  Richards B, W. E., Gabor D. Electromagnetic diffraction in optical systems, II. Structure of the image field in an aplanatic system. *Proc R Soc Lond Ser A: Math Phys Sci* **253**, 358-379, doi:10.1098/rspa.1959.0200 (1959).
3  Gao, J. *et al.* Polarization-conversion microscopy for imaging the vectorial polarization distribution in focused light. *Optica* **8**, doi:10.1364/optica.422836 (2021).
4  Wilson, T. & Sheppard, C. J. R. *Theory and Practice of Scanning Optical Microscopy*. (Academic Press, 1984).